\documentclass[11pt,onecolumn,draftcls]{IEEEtran}
\usepackage{graphics}
\usepackage{graphicx}
\usepackage{epsfig,amssymb}
\usepackage{amsmath}
\usepackage{times}
\usepackage{mathrsfs}
\usepackage{array}
\usepackage{amsmath, latexsym, amsfonts, amssymb}
\usepackage[mathscr]{eucal}
\usepackage{array, tabularx}
\usepackage[table]{xcolor}
\usepackage{psfrag}
\usepackage{multirow}
\usepackage{booktabs}
\usepackage{epsfig}
\usepackage{cite}
\usepackage{tikz}

\newcommand{\paren}[1]{\left(#1\right)}
\newcommand{\sqparen}[1]{\left[#1\right]}
\newcommand{\brparen}[1]{\left\{#1\right\}}
\newcommand{\field}[1]{\ensuremath{\mathbb{#1}}}
\newcommand{\R}{\ensuremath{\field{R}}} 
\newcommand{\Rp}{\ensuremath{\R_+}} 
\newcommand{\I}[1]{\ensuremath{\mathsf{1}_{\left\{#1\right\}}}} 
\newcommand{\PR}[1]{\ensuremath{\mathsf{Pr}\left\{#1\right\}}} 
\newcommand{\ES}[1]{\ensuremath{\mathsf{E}\left[#1 \right]}} 
\newcommand{\sinr}{\ensuremath{{\rm SINR}}}

\newcommand{\imax}[2]{\ensuremath{#1}_{\ensuremath{{\paren{#2}}}}}
\newcommand{\major}{\ensuremath{\succeq_{\rm M}}}

\renewcommand{\vec}[1]{\ensuremath{\boldsymbol{#1}}} 

\newtheorem{theorem}{Theorem}
\newtheorem{lemma}{Lemma}
\newtheorem{definition}{Definition}

\begin{document}
\title{Vector Broadcast Channels: Optimal Threshold Selection Problem}
\author{\authorblockN{Tharaka Samarasinghe, Hazer Inaltekin and Jamie Evans\\}
\authorblockA{Department of Electrical and Electronic Engineering, The University of Melbourne, Australia.\\ Email: \{ts, hazeri, jse\}@unimelb.edu.au}}
\date{}
\bibliographystyle{ieeetr}
\maketitle

\begin{abstract}
Threshold feedback policies are well known and provably rate-wise optimal selective feedback techniques for communication systems requiring partial channel state information (CSI).  However, optimal selection of thresholds at mobile users to maximize information theoretic data rates subject to feedback constraints is an open problem.  In this paper, we focus on the optimal threshold selection problem, and provide a solution for this problem for finite feedback systems.  Rather surprisingly, we show that using the same threshold values at all mobile users is not always a rate-wise optimal feedback strategy, even for a system with identical users experiencing statistically the same channel conditions.  By utilizing the theory of majorization, we identify an underlying Schur-concave structure in the rate function and obtain sufficient conditions for a homogenous threshold feedback policy to be optimal.  Our results hold for most fading channel models, and we illustrate an application of our results to familiar Rayleigh fading channels.
\end{abstract}

\section{Introduction}
Consider a multiuser communication system in a fading environment. The channel changes over time, and the goal of the base station (BS) is to maximize downlink data rates by taking channel variations into account.  The BS has multiple transmission antennas, and therefore can possibly communicate with a selected subset of users through multiple traffic flows simultaneously.  In this setting, selecting the users with the best instantaneous signal-to-interference-plus-noise-ratios ($\sinr$) for communication is a simple communication strategy that is heuristically expected to maximize downlink data rates.  Indeed, this is the classical opportunistic communication approach utilizing multiuser diversity to take advantage of changing channel conditions for rate maximization \cite{viswanathtse, Tse02}.

For vector broadcast channels, it is well known that such an opportunistic communication strategy is asymptotically optimal \cite{hassibi}.  Furthermore, this approach only requires mobile users to feed back their partial CSI (i.e., their $\sinr$s), as opposed to other approaches requiring perfect CSI \cite{shamai03, VJG03}.  However, regardless of partial or perfect CSI feedback, all users contend for the uplink to communicate their CSI to the BS in such a setting.  This is an impractical burden on the uplink (for large numbers of users), and a significant waste of communication resources, which is created by the feedback from users having no realistic chance of being scheduled for communication. These considerations motivate the use of selective feedback techniques in which only the users with channels good enough are allowed to feed back \cite{Gesbert04}.  As shown in our companion paper \cite{Tharaka-ISIT1}, threshold feedback policies \cite{Diaz06, Pugh10, Hazer-WiOpt, Tharaka-ICC} form a rate-wise optimal subset of selective feedback policies for downlink throughput maximization subject to constraints on the average number of users feeding back.  According to a threshold feedback policy, each user decides to feed back or not by comparing a metric that captures channel quality (mostly $\sinr$ values) with her threshold value.  Different users are allowed to have different thresholds if such an heterogeneity in thresholds maximizes downlink throughput.  This paper studies the important issue of how to set threshold values at mobile users optimally to maximize downlink data rates without violating finite feedback constraints.

To this end, we focus on the general setting of vector broadcast channels operating based on the concept of opportunistic beamforming \cite{hassibi}: Each user calculates the $\sinr$ on each beam (random orthonormal beams in this context) and feeds back the $\sinr$ value and the index of the best beam to the BS.  Upon retrieval of this information, the BS schedules the user with the best $\sinr$ on each beam.  Certainly, threshold feedback policies provide further feedback reductions for such systems, and therefore the following question is of practical and theoretical importance to answer: What is the optimal assignment of thresholds to users so that aggregate data rates over multiple beams are maximized subject to constraints on the average number of users feeding back per beam?

Recently, a rather surprising answer for this question was presented in \cite{Hazer-WiOpt}.  For a system with identical users experiencing the same channel conditions statistically, it is intuitively expected that using the same threshold value at all users is rate-wise optimal.  However, Inaltekin et. al. in \cite{Hazer-WiOpt} showed that this is not always true, and established sufficient conditions on $\sinr$ distributions for the optimality of homogenous threshold feedback policies.  The limitation of \cite{Hazer-WiOpt} is that they only focused on two-user communication systems.  In this paper, we use the theory of majorization \cite{Marshall79} to extend this result to communication systems containing arbitrary number of users. 

The rest of the paper is organized as follows.  We formally define threshold feedback policies, and formulate the optimal threshold selection problem in Section \ref{Section: System Model}.  Section \ref{Section: Majorization} provides some key concepts from the theory of majorization to be utilized in later parts of the paper.  The aggregate rate function is analyzed (as a function of threshold values and feedback probabilities) in Section \ref{Section: Analysis of Schur-concavity}.  This analysis leads to sufficient conditions on the $\sinr$ distributions to ensure Schur-concavity of the rate function, and thereby the optimality of homogenous threshold feedback policies.  We illustrate an application of our results
to familiar Rayleigh fading channels in Section \ref{Section: Applications and Discussion}, and Section
\ref{Conclusions} concludes the paper.

\section{System Model} \label{Section: System Model}
Consider a  vector broadcast channel with $M$ beams.  There are $n$ users with each having a single receiver antenna.  The $\sinr$ on beam $m$ at user $i$ is given by
\begin{eqnarray}
\gamma_{i, m} =  \frac{\left| X_{i, m} \right|^2}{\frac{1}{\rho} +
\sum_{k=1, k \neq m}^M \left| X_{i, k} \right|^2}, \label{Eqn:
SINR Expression}
\end{eqnarray}
where $\left| X_{i, k} \right|$ is the received instantaneous signal power on beam $k$ at user $i$, and $\frac{1}{\rho}$ is the background noise power.  The random received signal powers at different users are assumed to be independent and identically distributed (i.i.d.).  This implies i.i.d. fading processes at different users.  Let $\vec{\gamma}_i = \paren{\gamma_{i, 1}, \gamma_{i, 2}, \cdots, \gamma_{i, M}}^\top \in \Rp^M$ be the $\sinr$ vector at user $i$.  Beams are identical.  That is,  for all $i \in \mathcal{N}$, the elements of $\vec{\gamma}_i$ are identically distributed (but dependent) with a common marginal distribution $F$, where $\mathcal{N}= \brparen{1,2,\ldots,n}$.  We will assume that $F$ is continuous, and has the density $f$ with support $\Rp$, which are true for many fading models.  Users feed back according to a threshold feedback policy. 
We formally define a threshold feedback policy as follows.

\begin{definition}\label{thresh policy}
We say $\vec{\mathcal{T}} = \paren{\mathcal{T}_1, \mathcal{T}_2, \cdots, \mathcal{T}_n}^\top$ is a {\em threshold feedback policy} if, for all $i \in \mathcal{N}$, there is a threshold $\tau_i$ such that $\mathcal{T}_i\paren{\vec{\gamma}_i}$ generates a feedback packet containing $\sinr$ values $\brparen{\gamma_{i, k}}_{k \in \mathcal{I}_i}$ if and only if $\gamma_{i, k} \geq \tau_i$ for all $k \in \mathcal{I}_i \subseteq \mathcal{M}$, where $\mathcal{M} = \brparen{1, 2, \cdots, M}$.  We call it a {\em homogenous threshold feedback policy} if all users use the same threshold $\tau$, i.e., $\tau_i =\tau$ for all $i \in \mathcal{N}$.
\end{definition}

Upon retrieval of feedback information, the BS schedules communication to the user with the highest $\sinr$ on each beam.  We define the truncated $\sinr$ on beam $m$ at user $i$ as $\bar{\gamma}_{i, m}=\gamma_{i,m}\I{\gamma_{i,m}\geq \tau_i}$, and let $\vec{\bar{\gamma}}_i$ be the truncated $\sinr$ vector at user $i$.  This vector contains all useful $\sinr$ values to be fed back by user $i$ on a positive feedback decision.  Then, the ergodic sum rate achieved for a given vector of thresholds $\vec{\tau}$ is given by
\begin{eqnarray}
R\paren{\vec{\tau}} =
\ES{r\paren{\brparen{\vec{\bar{\gamma}}_i}_{i=1}^n}}
=\ES{\sum_{m=1}^M \log\paren{1 +
\max_{1 \leq i \leq n} \bar{\gamma}_{i, m}}}, 
\nonumber \label{Eqn: Rate Def}
\end{eqnarray}
where $r\paren{\brparen{\vec{\bar{\gamma}}_i}_{i=1}^n}$ is the instantaneous downlink communication rate.  Note that zero rate is achieved on a beam if no user requests it. Conditioned on an event $\mathcal{A}$ (or, a random variable), we define the conditional rate as $R\paren{\vec{\tau}|\mathcal{A}}= \ES{r\paren{\brparen{\vec{\bar{\gamma}}_i}_{i=1}^n}|\mathcal{A}}$.  We also let $R^m\paren{\vec{\tau}}$ and $r^m\paren{\brparen{\vec{\bar{\gamma}}_i}_{i=1}^n}$ be the ergodic sum rate and the instantaneous rate on beam $m$, respectively.

$\Lambda\paren{\vec{\tau}}$ denotes the average number of users feeding back per beam.  It measures the performance of the system along the feedback dimension.  We have $\Lambda(\vec{\tau}) = \sum_{i=1}^n \PR{\gamma_{i,1} \geq \tau_i}$ since beams are statistically identical.  We let $p_i = \PR{\gamma_{i,1} \geq \tau_i}$.  Given a finite feedback system with a feedback constraint $\lambda$, the main optimization problem to be solved is to determine the optimal threshold values  to maximize the downlink throughput, which is formally written as
\begin{eqnarray}
\begin{array}{ll}
\underset{\vec{\tau} \in \Rp^n}{\mbox{maximize}} & R\paren{\vec{\tau}} \\
\mbox{subject to} & \sum_{i = 1}^n \PR{\gamma_{i, 1} \geq \tau_i}
\leq \lambda
\end{array}. \label{Optimization Problem}
\end{eqnarray}

As shown in \cite{Hazer-WiOpt}, this optimization problem is not easy to solve even for a two-user system due to the non-convex objective function and the non-convex constraint set depending on the distribution of the $\sinr$ values.  The complexity of the problem increases with increasing numbers of users due to the dimensionality growth.  In this paper, we will use the theory of majorization \cite{Marshall79} to analyze the underlying Schur-concave structure in the objective rate function, and solve this optimization problem for a general $n$-user system. Schur-convex/concave functions frequently arise in mathematical analysis and engineering applications \cite{Arnold07}.  For example, every function that is convex and symmetric is also a Schur-convex function.  In the next section, we will briefly introduce some key concepts from the theory of majorization to be used in our analysis later.

\section{Majorization}\label{Section: Majorization}
For a vector $\vec{x}$ in $\R^n$, we denote its ordered coordinates by $\imax{x}{1} \geq \cdots \geq \imax{x}{n}$.  For $\vec{x}$ and $\vec{y}$ in $\R^n$, we say $\vec{x}$ {\em majorizes} $\vec{y}$ and write it as $\vec{x} \major \vec{y}$ if we have $\sum_{i=1}^j \imax{x}{i} \geq \sum_{i=1}^j \imax{y}{i}$ when $j = 1, \cdots, n-1$, and $\sum_{i=1}^n \imax{x}{i} = \sum_{i=1}^n \imax{y}{i}$.  A function $\varphi: \R^n \mapsto \R$ is said to be {\em Schur-convex} if $\vec{x} \major \vec{y}$ implies $\varphi\paren{\vec{x}} \geq \varphi\paren{\vec{y}}$, and $\varphi$ is Schur-concave if $-\varphi$ is Schur-convex.  A Schur-concave function tends to increase when the components of its argument become more similar.  We will establish conditions under which $R\paren{\vec{\tau}}$ becomes a Schur-concave function, which will, in turn, imply the optimality of homogenous threshold feedback policies.

However, establishing these conditions is not straightforward, which makes the following lemma vital for our analysis.
\begin{lemma}\label{lemma:majorization}
Let $\varphi$ be a real-valued function defined on $\Rp^n$, and $\mathcal{D}= \brparen{\vec{z} \in \Rp^n: z_1 \geq z_2 \geq \cdots \geq z_n}$.  Then, $\varphi$ is a Schur-convex function if and only if, for all $\vec{z} \in \mathcal{D}$ and $i=1, 2, \cdots,n-1$, $\varphi\paren{z_1,\ldots,z_{i-1},z_i+\epsilon,z_{i+1}-\epsilon, z_{i+2}, \ldots,z_{n}}$ is increasing in $\epsilon$ over the region $0 \leq \epsilon \leq \min\paren{z_{i-1}-z_{i},z_{i+1}-z_{i+2}}$. \footnote{At the end points $i=1$, $i = n-1$, the condition is modified accordingly.}
\end{lemma}

It can be seen that the coordinates $z_i$ and $z_{i+1}$ are systematically altered by using the parameter $\epsilon$, and the constraints on $\epsilon$ eliminate any violation in the order.  Interested readers are referred to \cite{Marshall79} for more insights on the theory of majorization.  In the next section, we will see how we can use this theory to identify the Schur-concave structure in the objective rate function.

\section{Schur-concavity Analysis}\label{Section: Analysis of Schur-concavity}
In this section, we will establish sufficient conditions on the $\sinr$ distributions for the Schur-concavity of the rate function.  We focus on the first beam to explain our proof ideas without any loss of generality since all beams are statistically identical. We start our analysis by analyzing the rate as a function of thresholds and establishing three important lemmas.  Then, we will incorporate the feedback constraint into our optimization problem by interpreting the rate as a function of feedback probabilities.  Using these results, we will finally establish the underlying Schur-concave structure in the rate function through the theory of majorization.

\subsection{Rate as a Function of Thresholds}

Consider thresholds in increasing order, i.e., $\tau_{\pi(1)} \leq \cdots \leq \tau_{\pi(i)} \leq \tau_{\pi(i+1)}\cdots \leq \tau_{\pi(n)}$.  Now, by analyzing $ R^1\paren{\tau_{\pi(i+1)} +\epsilon,\tau_{\pi(i)}-\epsilon} = R^1\paren{\tau_{\pi(n)} , \cdots, \tau_{\pi(i+1)} +\epsilon,\tau_{\pi(i)}-\epsilon ,\cdots , \tau_{\pi(1)} }$, we can use Lemma \ref{lemma:majorization} to identify the underlying Schur-concave structure in the rate function.  However, analysis of this function is still complex. Therefore, we use the following method to reduce problem complexity.

Let $\mathcal{N}^\prime=\brparen{k\  :\ k \in \mathcal{N} \ \& \ k \neq \pi(i),\pi(i+1)}$.  We fix the thresholds and the $\sinr$ values of all users in $\mathcal{N}^\prime$.  Randomness is now associated only with users $\pi(i)$ and $\pi(i+1)$.  Let $\bar{\gamma}^\star_{\mathcal{N}^\prime}=\underset{k \in \mathcal{N}^\prime}{\max}\gamma_{k,1}\I{\gamma_{k, 1} \geq \tau_{k}}$.  The instantaneous rate on beam $1$ as a function of $\bar{\gamma}_{\pi(i),1}$, $\bar{\gamma}_{\pi(i+1),1}$ and $\bar{\gamma}^\star_{\mathcal{N}^\prime}$ is $r^1\paren{\bar{\gamma}_{\pi(i),1},\bar{\gamma}_{\pi(i+1),1},\bar{\gamma}^\star_{\mathcal{N}^\prime}} = \log\paren{1+
\max\brparen{\bar{\gamma}_{\pi(i),1},\bar{\gamma}_{\pi(i+1),1},\bar{\gamma}^\star_{\mathcal{N}^\prime}}}$.
Therefore,
\begin{eqnarray}\label{eqn:conditional expec}
R^1\paren{\tau_{\pi(i)},\tau_{\pi(i+1)}|\bar{\gamma}^\star_{\mathcal{N}^\prime}}=
\ES{r^1\paren{\bar{\gamma}_{\pi(i)},\bar{\gamma}_{\pi(i+1)},\bar{\gamma}^\star_{\mathcal{N}^\prime}}|\bar{\gamma}^\star_{\mathcal{N}^\prime}}.
\end{eqnarray}

It is not hard to see that $R^1\paren{\tau_{\pi(i)},\tau_{\pi(i+1)}|\bar{\gamma}^\star_{\mathcal{N}^\prime}}$ depends on the value of $\bar{\gamma}^\star_{\mathcal{N}^\prime}$.  $\bar{\gamma}^\star_{\mathcal{N}^\prime}$ can be greater than $\tau_{\pi(i+1)}$, less than $\tau_{\pi(i)}$ or in between $\tau_{\pi(i)}$ and $\tau_{\pi(i+1)}$.  We will now establish three important lemmas for these three cases, which will be useful in interpreting the rate. We will start with the case $\bar{\gamma}^\star_{\mathcal{N}^\prime} >\tau_{\pi(i+1)}$.

\begin{lemma}\label{lemma:greater than thresh}
If $\bar{\gamma}^\star_{\mathcal{N}^\prime} >\tau_{\pi(i+1)}$,
$R^1\paren{\tau_{\pi(i)},\tau_{\pi(i+1)}|\bar{\gamma}^\star_{\mathcal{N}^\prime}}$
only depends on $\bar{\gamma}^\star_{\mathcal{N}^\prime}$, and
given by $G(\bar{\gamma}^\star_{\mathcal{N}^\prime})$, where
\begin{multline*}
G(\bar{\gamma}^\star_{\mathcal{N}^\prime})=
\PR{\gamma^\star_{i,i+1}\leq\bar{\gamma}^\star_{\mathcal{N}^\prime}|\bar{\gamma}^\star_{\mathcal{N}^\prime}
}\log\paren{1+\bar{\gamma}^\star_{\mathcal{N}^\prime}} \\+
\ES{\log\paren{1 +
\gamma^\star_{i,i+1}}\I{\gamma^\star_{i,i+1}>\bar{\gamma}^\star_{\mathcal{N}^\prime}}|\bar{\gamma}^\star_{\mathcal{N}^\prime}},
\end{multline*}
and
$\gamma^\star_{i,i+1}=\max\brparen{\gamma_{\pi(i)},\gamma_{\pi(i+1)}}$.
\end{lemma}
\begin{IEEEproof}
Let
$\bar{\gamma}^\star_{i,i+1}=\max\brparen{\bar{\gamma}_{\pi(i)},\bar{\gamma}_{\pi(i+1)}}$.
From (\ref{eqn:conditional expec}),
\begin{multline}\label{eqn:conditional expec1}
R^1\paren{\tau_{\pi(i)},\tau_{\pi(i+1)}|\bar{\gamma}^\star_{\mathcal{N}^\prime}}
\\=
\log\paren{1+\bar{\gamma}^\star_{\mathcal{N}^\prime}}\PR{
\bar{\gamma}^\star_{i,i+1}
\leq \bar{\gamma}^\star_{\mathcal{N}^\prime}|\bar{\gamma}^\star_{\mathcal{N}^\prime} } \\
+ \ES{\log\paren{1 +
\bar{\gamma}^\star_{i,i+1}}\I{\bar{\gamma}^\star_{i,i+1}>\bar{\gamma}^\star_{\mathcal{N}^\prime}}|\bar{\gamma}^\star_{\mathcal{N}^\prime}}.
\end{multline}

For the first term on the right hand side of \eqref{eqn:conditional expec1}, it is not hard to show that $\brparen{\bar{\gamma}^\star_{i,i+1} \leq \bar{\gamma}^\star_{\mathcal{N}^\prime}}= \brparen{\gamma^\star_{i,i+1} \leq \bar{\gamma}^\star_{\mathcal{N}^\prime}}$ since $\bar{\gamma}^\star_{\mathcal{N}^\prime} >\tau_{\pi(i+1)}$.

For the second term, we must have $\bar{\gamma}^\star_{i,i+1}=\gamma^\star_{i,i+1}$ since $\bar{\gamma}^\star_{i,i+1}>\bar{\gamma}^\star_{\mathcal{N}^\prime} > \tau_{\pi(i+1)}$, . Thus, we end up
with $R^1\paren{\tau_{\pi(i)},\tau_{\pi(i+1)}|\bar{\gamma}^\star_{\mathcal{N}^\prime}} = G(\bar{\gamma}^\star_{\mathcal{N}^\prime})$.
\end{IEEEproof}

The next lemma will evaluate the value of
$R^1\paren{\tau_{\pi(i)},\tau_{\pi(i+1)}|\bar{\gamma}^\star_{\mathcal{N}^\prime}}$
for $\bar{\gamma}^\star_{\mathcal{N}^\prime} <\tau_{\pi(i)} $.
\begin{lemma} \label{lamma2}\label{lemma:less than thresh}
If $\bar{\gamma}^\star_{\mathcal{N}^\prime} <\tau_{\pi(i)} $,
$R^1\paren{\tau_{\pi(i)},\tau_{\pi(i+1)}|\bar{\gamma}^\star_{\mathcal{N}^\prime}}$
is given by
\begin{multline*}
R_1\paren{\tau_{\pi(i)},\tau_{\pi(i+1)}|\bar{\gamma}^\star_{\mathcal{N}^\prime}}
=\int_{\tau_{\pi(i+1)}}^\infty
\log(1+x)dF^2(x)\\+F\paren{\tau_{\pi(i+1)}}\int_{\tau_{\pi(i)}}^{\tau_{\pi(i+1)}}
\log(1+x)dF(x)\\+
\log\paren{1+\bar{\gamma}^\star_{\mathcal{N}^\prime}}F\paren{\tau_{\pi(i)}}F\paren{\tau_{\pi(i+1)}}.
\end{multline*}
\end{lemma}
\begin{IEEEproof}
When $\bar{\gamma}^\star_{\mathcal{N}^\prime} <\tau_{\pi(i)}$,
(\ref{eqn:conditional expec1}) simplifies to
$R^1\paren{\tau_{\pi(i)},\tau_{\pi(i+1)}|\bar{\gamma}^\star_{\mathcal{N}^\prime}}=$
\begin{multline*}
\ES{\log\paren{1 +
\gamma^\star_{i,i+1}}\I{\gamma_{\pi(i),1}>\tau_{\pi(i)},\gamma_{\pi(i+1),1}>\tau_{\pi(i+1)}}}
\\+ \ES{\log\paren{1 +
\gamma_{\pi(i+1),1}}\I{\gamma_{\pi(i+1),1}>\tau_{\pi(i+1)}}}F\paren{\tau_{\pi(i)}}
\\+ \ES{\log\paren{1 +
\gamma_{\pi(i),1}}\I{\gamma_{\pi(i),1}>\tau_{\pi(i)}}}F\paren{\tau_{\pi(i+1)}}
\\+\log\paren{1+\bar{\gamma}^\star_{\mathcal{N}^\prime}}F\paren{\tau_{\pi(i)}}F\paren{\tau_{\pi(i+1)}}.
\end{multline*}

The first three terms on the right hand side is identical to the rate expression of the two-user system analyzed in \cite{Hazer-WiOpt}.  Substituting the result for the two-user case completes the proof.
\end{IEEEproof}

Finally, we look at the case where $\tau_{\pi(i)} \leq \bar{\gamma}^\star_{\mathcal{N}^\prime} \leq \tau_{\pi(i+1)} $.
\begin{lemma}\label{lemma3}\label{lemma:between the thresh}
If $\tau_{\pi(i)} \leq \bar{\gamma}^\star_{\mathcal{N}^\prime} \leq \tau_{\pi(i+1)} $, then the value of $R^1\paren{\tau_{\pi(i)},\tau_{\pi(i+1)}|\bar{\gamma}^\star_{\mathcal{N}^\prime}}$ is given by
\begin{multline*}
R_2\paren{\tau_{\pi(i)},\tau_{\pi(i+1)}|\bar{\gamma}^\star_{\mathcal{N}^\prime}}
= \int_{\tau_{\pi(i+1)}}^\infty \log(1+x)dF^2(x)
\\+F\paren{\tau_{\pi(i+1)}}\int_{\bar{\gamma}^\star_{\mathcal{N}^\prime}}^{\tau_{\pi(i+1)}}
\log(1+x)dF(x)\\+\log\paren{1+\bar{\gamma}^\star_{\mathcal{N}^\prime}}
F\paren{\tau_{\pi(i+1)}}
\paren{1-\PR{\gamma_{\pi(i)}\geq\bar{\gamma}^\star_{\mathcal{N}^\prime}|\bar{\gamma}^\star_{\mathcal{N}^\prime}
}}.
\end{multline*}
\end{lemma}

This lemma can be proved by using similar arguments to the ones used in the previous two lemmas.  We skip its proof to avoid repetitions.

If $\bar{\gamma}^\star_{\mathcal{N}^\prime}=\tau_{\pi(i)}$, $R_1$ and $R_2$ in Lemmas \ref{lemma:less than thresh} and \ref{lemma:between the thresh} evaluate to the same expression.  Similarly, if $\bar{\gamma}^\star_{\mathcal{N}^\prime}=\tau_{\pi(i+1)}$, $G$ and $R_2$ in Lemmas \ref{lemma:greater than thresh} and \ref{lemma:between the thresh} evaluate to the same expression.  This shows that the rate as a function of $\bar{\gamma}^\star_{\mathcal{N}^\prime}$ is continuous at $\tau_{\pi(i)}$ and $\tau_{\pi(i+1)}$. 

Given the initial threshold values $\brparen{\tau_{\pi(j)}}_{j=1}^n$, the first approach to discover the Schur-concave structure in the rate function is to analyze the behavior of the function $$g_T(\epsilon)=R^1\paren{\tau_{\pi(i)}-\epsilon,\tau_{\pi(i+1)}+\epsilon|\bar{\gamma}^\star_{\mathcal{N}^\prime}}$$ for $\epsilon \in \sqparen{0, \min\paren{\tau_{\pi(i)}-\tau_{\pi(i-1)},\tau_{\pi(i+2)}-\tau_{\pi(i+1)}}}$ by making use of Lemma \ref{lemma:majorization}.  This is now a scalar problem.  However, we find it more useful to interpret the rate as a function of feedback probabilities since the feedback constraint in \eqref{Optimization Problem} is in terms of these probabilities. Through this interpretation, we can incorporate the feedback constraint into our optimization problem more easily, as will be shown in the next sub-section.

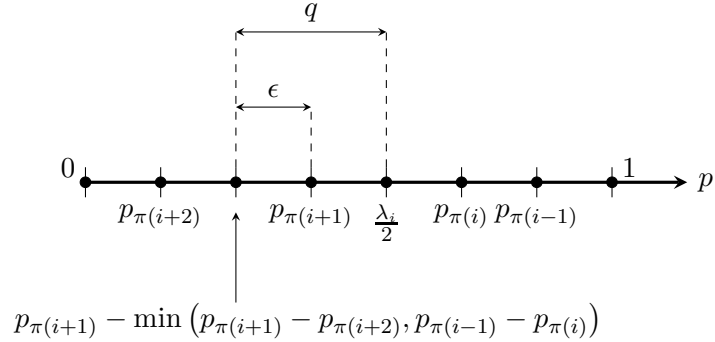
\begin{figure}[t]
\begin{center}
\hspace{0cm}
\begin{tikzpicture}[
        scale=2]
    \draw[->, very thick, black, >=stealth] (0, 0) -- (4.0, 0) node[right]{$p$};
     \filldraw[fill=black] (3.5, 0) circle (1pt);
     \draw[thin] (3.5, -0.1) -- (3.5, 0.1) node[right]{$1$};
     \filldraw[fill=black] (0, 0) circle (1pt);
      \draw[thin] (0, -0.1) -- (0, 0.1) node[left]{$0$};
      \draw[thin] (0.5, -0.1) -- (0.5, 0.1) node[pos = 0.1, below]{$p_{\pi(i+2)}$};
      \filldraw[fill=black] (0.5, 0) circle (1pt);
      \draw[thin] (1, -0.1) -- (1, 0.1);
      \draw[->, thin, black, >=stealth] (1, -0.8) -- (1, -0.2) node[pos = -0.2, left=-5cm]{$\hspace{0cm} p_{\pi(i+1)} - \min\left(p_{\pi(i+1)} - p_{\pi(i+2)}, p_{\pi(i-1)} - p_{\pi(i)} \right)$};
      \filldraw[fill=black] (1, 0) circle (1pt);
      \draw[thin, dashed] (1,0.1) -- (1, 1);
      \draw[thin] (1.5, -0.1) -- (1.5, 0.1) node[pos = 0.1, below]{$p_{\pi(i+1)}$};
       \filldraw[fill=black] (1.5, 0) circle (1pt);
       \draw[thin, dashed] (1.5, 0.1) -- (1.5, 0.5);
       \draw[<->, thin, black, >=stealth] (1, 0.5) -- (1.5, 0.5) node[above, midway]{$\epsilon$};
       \draw[thin] (2.0, -0.1) -- (2.0, 0.1) node[pos = 0.1, below]{$\frac{\lambda_i}{2}$};
       \filldraw[fill=black] (2.0, 0) circle (1pt);
       \draw[thin, dashed] (2.0, 0.1) -- (2.0, 1);
        \draw[<->, thin, black, >=stealth] (1, 1) -- (2, 1) node[above, midway]{$q$};
       \draw[thin] (2.5, -0.1) -- (2.5, 0.1) node[pos = 0.1, below]{$p_{\pi(i)}$};
        \filldraw[fill=black] (2.5, 0) circle (1pt);
        \draw[thin] (3.0, -0.1) -- (3.0, 0.1) node[pos = 0.1, below]{$p_{\pi(i-1)}$};
        \filldraw[fill=black] (3.0, 0) circle (1pt);
\end{tikzpicture}
\end{center}
\vspace{-0.3cm}
\caption{Ordered feedback probabilities, and the range of $q$ and $\epsilon$.} \label{line}
\end{figure}
\vspace{-0.75cm}

\subsection{Rate as a Function of Feedback Probabilities}


Since $\tau_{\pi(i)}=F^{-1}(1-p_{\pi(i)} )$ and $f$ has the support $\Rp$, we can write $R^1\paren{\tau_{\pi(i)},\tau_{\pi(i+1)}|\bar{\gamma}^\star_{\mathcal{N}^\prime}}$ as $R^1\paren{p_{\pi(i)},p_{\pi(i+1)}|\bar{\gamma}^\star_{\mathcal{N}^\prime}}$ without any ambiguity. Since $F$ is monotone increasing, we have $p_{\pi(1)} \geq p_{\pi(2)} \geq \cdots \geq p_{\pi(i)} \geq p_{\pi(i+1)} \geq \cdots \geq p_{\pi(n)}$. Therefore, given the initial feedback probabilities $\brparen{p_{\pi(j)}}_{j=1}^n$, we analyze the behavior of the function
\begin{eqnarray}
g_p(\epsilon)=R^1\paren{p_{\pi(i)}+\epsilon,p_{\pi(i+1)}-\epsilon|\bar{\gamma}^\star_{\mathcal{N}^\prime}}
\end{eqnarray}
for $\epsilon \in \sqparen{0, \min\paren{p_{\pi(i-1)}-p_{\pi(i)},p_{\pi(i+1)}-p_{\pi(i+2)}}}$ to discover Schur-concavity of the rate function by Lemma \ref{lemma:majorization}.  $p_{\pi(i)}$ and $p_{\pi(i+1)}$ give us the feedback level $\lambda_i=p_{\pi(i)}+p_{\pi(i+1)}$, and other probabilities give us natural boundaries on $p_{\pi(i)}$ and  $p_{\pi(i+1)}$ as $p_{\pi(i+2)} \leq p_{\pi(i+1)} \leq p_{\pi(i)} \leq p_{\pi(i-1)}$.  Without violating these boundaries, we can vary $p_{\pi(i+1)}$ by keeping $\lambda_i$ constant.  That is, we can take any $q \in \mathcal{P}_{i+1}$ and write $R^1\paren{p_{\pi(i)}+\epsilon,p_{\pi(i+1)}-\epsilon|\bar{\gamma}^\star_{\mathcal{N}^\prime}}$ as a function of $q$, where $\mathcal{P}_{i+1} = \sqparen{p_{\pi(i+1)}-\min\paren{p_{\pi(i+1)}-p_{\pi(i+2)},p_{\pi(i-1)}-p_{\pi(i)}}, \frac{\lambda_i}{2}}$.  Please refer to Fig. \ref{line} for a graphical representation of this selection of $q$.


Now, by using Lemma \ref{lemma:greater than thresh}, \ref{lemma:less than thresh} and \ref{lemma:between the thresh}, we have
$R^1\paren{q | \bar{\gamma}^\star_{\mathcal{N}^\prime}} = G\paren{\bar{\gamma}^\star_{\mathcal{N}^\prime}}\I{q > 1- F\paren{\bar{\gamma}^\star_{\mathcal{N}^\prime}}}+ R_1\paren{q | \bar{\gamma}^\star_{\mathcal{N}^\prime}}\I{q > \lambda_i -\paren{1- F\paren{\bar{\gamma}^\star_{\mathcal{N}^\prime}}}}+ R_2\paren{q | \bar{\gamma}^\star_{\mathcal{N}^\prime}}\I{q \leq 1- F(\bar{\gamma}^\star_{\mathcal{N}^\prime})\  \&\ q \leq \lambda_i -\paren{1- F\paren{\bar{\gamma}^\star_{\mathcal{N}^\prime}}}}$
for all $q \in \mathcal{P}_{i+1}$.


Some insights about this rate expression are as follows.  Assume $1-F\paren{\bar{\gamma}^\star_{\mathcal{N}^\prime}} \leq \frac{\lambda_i}{2}$.  Then, when $q$ changes from $p_{\pi(i+1)}-\min\paren{p_{\pi(i+1)}-p_{\pi(i+2)},p_{\pi(i-1)}-p_{\pi(i)}}$ to $\frac{\lambda_i}{2}$, the threshold of the user with index $\pi(i+1)$ can become as small as $F^{-1}\paren{1-\frac{\lambda_i}{2}}$, and can become as large as $\tau_{\pi(i+2)}$.
Therefore, when $\tau_{\pi(i+1)}$ is large, $\bar{\gamma}^\star_{\mathcal{N}^\prime}$ lies in between two thresholds, and the rate expression is given by $R_2\paren{q | \bar{\gamma}^\star_{\mathcal{N}^\prime}}$ for small values of $q$.  When $\tau_{\pi(i+2)}$ is small, $\bar{\gamma}^\star_{\mathcal{N}^\prime}$ always lies to the right of $\tau_{\pi(i+1)}$, and the rate is given by the constant $G\paren{\bar{\gamma}^\star_{\mathcal{N}^\prime}}$.  Similar explanations can be given for the rate expression for $1- F\paren{\bar{\gamma}^\star_{\mathcal{N}^\prime}} > \frac{\lambda_i}{2}$.
\subsection{Schur-concavity of the Rate Function}
Now, we show how we can utilize the theory of majorization to identify the Schur-concavity of the rate function.  First, we show that $R^1\paren{q | \bar{\gamma}^\star_{\mathcal{N}^\prime}}$ is an increasing function of $q$ over $\mathcal{P}_{i+1}$.  Note that we are proving that the rate function is increasing over a larger region than the one needed in Lemma \ref{lemma:majorization}. (Please refer to Fig. \ref{line}.)  The following lemma formally states sufficient conditions for this property to be true.  
\begin{lemma}\label{lemma:increasing function}
$R^1\paren{q | \bar{\gamma}^\star_{\mathcal{N}^\prime}}$ is an increasing function of $q$ for any given
$\bar{\gamma}^\star_{\mathcal{N}^\prime}$ if $f$ is bounded at zero, and has the derivative $f^\prime$ satisfying $f^\prime\paren{F^{-1}(q)} \leq -\frac{f\paren{F^{-1}(q)}}{\paren{1+F^{-1}(q)}}$
for all $q \in \mathcal{P}_{i+1}$ and $i=1,\ldots,n-1$.
\end{lemma}
\begin{IEEEproof}
We can write $R_1\paren{q | \bar{\gamma}^\star_{\mathcal{N}^\prime}}$ and $R_2\paren{q | \bar{\gamma}^\star_{\mathcal{N}^\prime}}$ explicitly as
\begin{multline*}
R_1\paren{q | \bar{\gamma}^\star_{\mathcal{N}^\prime}}= \int_{F^{-1}(1-q)}^\infty \log(1+x)dF^2(x)\\+\paren{1-q}\int_{F^{-1}(1+q-\lambda_i)}^{F^{-1}(1-q)} \log(1+x)dF(x)\\+\log\paren{1+\bar{\gamma}^\star_{\mathcal{N}^\prime}}\paren{1-q}F\paren{1+q-\lambda_i},
\end{multline*}
and
\begin{multline*}
R_2\paren{q | \bar{\gamma}^\star_{\mathcal{N}^\prime}} = \int_{F^{-1}(1-q)}^\infty \log(1+x)dF^2(x)\\+\paren{1-q}\int_{\bar{\gamma}^\star_{\mathcal{N}^\prime}}^{F^{-1}(1-q)} \log(1+x)dF(x)\\+\log\paren{1+\bar{\gamma}^\star_{\mathcal{N}^\prime}}\paren{1-q}F\paren{\bar{\gamma}^\star_{\mathcal{N}^\prime}},
\end{multline*}
respectively. We can show that $ \frac{dR_2\paren{q | \bar{\gamma}^\star_{\mathcal{N}^\prime}}}{dq}\geq 0$
for any $q \leq \frac{\lambda_i}{2}$ by directly using the two-user analysis in \cite{Hazer-WiOpt}, given the conditions on the distribution are fulfilled.  Integration by parts on the derivative of $R_1\paren{q | \bar{\gamma}^\star_{\mathcal{N}^\prime}}$ gives us
$
\frac{dR_1\paren{q | \bar{\gamma}^\star_{\mathcal{N}^\prime}}}{dq}= \int_{\bar{\gamma}^\star_{\mathcal{N}^\prime}}^{F^{-1}(1-q)} \frac{F(x)}{1+x}dx \geq 0.
$
Since $G(\bar{\gamma}^\star_{\mathcal{N}^\prime})$ is independent of $q$, $\frac{dR^1\paren{q | \bar{\gamma}^\star_{\mathcal{N}^\prime}}}{dq}  \geq 0$ over $ \mathcal{P}_{i+1}$, which completes the proof.
\end{IEEEproof}

The above lemma proves that if the conditions on the $\sinr$ distribution are satisfied, $R^1\paren{q | \bar{\gamma}^\star_{\mathcal{N}^\prime}}$ is an increasing function of $q$.  This implies that $R^1\paren{p_{\pi(i)}+\epsilon, p_{\pi(i+1)}-\epsilon|\bar{\gamma}^\star_{\mathcal{N}^\prime}}$ is a decreasing function over the region of interest for $\epsilon$. (Please refer to Fig. \ref{line}.) Therefore, we can claim that the rate is a Schur-concave function of the feedback probabilities if the conditions in Lemma \ref{lemma:increasing function} are satisfied.  This is formally stated and proven in the next theorem.
\begin{theorem}\label{Thm: Shur-concavity of Rate}
$R\paren{\vec{p}}$ is Schur-concave if $f$ is bounded at zero, and has the derivative $f^\prime$ satisfying $f^\prime\paren{F^{-1}(x)} \leq -\frac{f\paren{F^{-1}(x)}}{\paren{1+F^{-1}(x)}}$ for all $x \in [0,1]$.
\end{theorem}
\begin{IEEEproof}
By Lemma \ref{lemma:increasing function}, $-R^1\paren{p_{\pi(i)}+\epsilon,p_{\pi(i+1)}-\epsilon|\bar{\gamma}^\star_{\mathcal{N}^\prime}}$ is an increasing function for any given $\bar{\gamma}^\star_{\mathcal{N}^\prime}$ over $\epsilon \in \sqparen{0, \min\paren{p_{\pi(i-1)}-p_{\pi(i)},p_{\pi(i+1)}-p_{\pi(i+2)}}}$. 
Therefore, by using Lemma \ref{lemma:majorization} and taking the expectation over $\bar{\gamma}^\star_{\mathcal{N}^\prime}$,  we have 
\begin{eqnarray*}
-R^1\paren{p_{\pi(i)}+\epsilon,p_{\pi(i+1)}-\epsilon} \geq -R^1\paren{p_{\pi(i)},p_{\pi(i+1)}}
\end{eqnarray*}
for all $\epsilon \in \sqparen{0, \min\paren{p_{\pi(i-1)}-p_{\pi(i)},p_{\pi(i+1)}-p_{\pi(i+2)}}}$.
Since this is correct for all $\vec{p} \in [0, 1]^n$, $-R^1\paren{\vec{p}}$ is a Schur-convex function of $\vec{p}$, implying that $R^1\paren{\vec{p}}$ is Schur-concave.  Since beams are identical, we also conclude that $R\paren{\vec{p}}$ is a Schur-concave function of feedback probabilities if the conditions in the theorem are met.
\end{IEEEproof}

A Schur-concave function increases when the dispersion among the components of its argument decreases. Therefore, this theorem implies that a solution for the optimization problem in (\ref{Optimization Problem})
is $\vec{\tau}^\star = \paren{F^{-1}\paren{1- \frac{\lambda}{N}}, F^{-1}\paren{1- \frac{\lambda}{N}}, \cdots, F^{-1}\paren{1- \frac{\lambda}{N}}}^\top$ if $f^\prime\paren{F^{-1}(x)} \leq -\frac{f\paren{F^{-1}(x)}}{\paren{1+F^{-1}(x)}}$ for all $x \in [0,1]$.  Next, we will see how our results can be applied to a Rayleigh fading channel model, which is one of the well known channel models in the literature, e.g., \cite{BPS98, Sklar97a}, and closely approximates measured data rates in densely populated urban areas \cite{Chizhik03}.

\section{Applications: Rayleigh Fading Channels} \label{Section: Applications and Discussion}
In this example, we will assume that transmitted signals and Rayleigh distributed channel fading coefficients are of unit power.  Interested readers are referred to \cite{hassibi} for further details of the physical layer model.  In this case, $F$ and $f$ can be given as $F(x)=1-\frac{e^{-\frac{x}{\rho}}}{(x+1)^{M-1}}$ and $f(x) = \frac{e^{-\frac{x}{\rho}}}{(x+1)^{M}}\sqparen{\frac{1}{\rho}(x+1)+M-1}$, and the functional inverse of $F$, $F^{-1}$, is given as $F^{-1}(x)=-\rho\log{(1-x)}$ if $M=1$ and $F^{-1}(x) = -1+(M-1)\rho W\paren{\frac{\exp\paren{\frac{1}{(M-1)\rho}}}{(M-1)\rho}(1-x)^{\frac{1}{1-M}}}$ if $M \geq 2$,
where $x \in [0, 1]$ and $W$ is the Lambert W function given by the defining equation $W(x)\exp(W(x))=x$ for $x \geq -\frac{1}{e}$.

We summarize our findings for Rayleigh fading environments as follows.
\begin{theorem} \label{Thm: Shur-concavity of Rate - SB}
For a Rayleigh fading environment with $M=1$, $R\paren{\vec{p}}$ is a Schur-concave function of feedback probabilities, and therefore a homogenous threshold feedback policy solves \eqref{Optimization Problem} if $\rho \leq 1$.
\end{theorem}

According to this theorem, a homogenous threshold feedback policy will be optimal if the background noise power is large enough,  i.e., signal-to-noise-ratio less than $0$dB. Although, intuitively, a homogenous threshold feedback policy may sound optimal for all channel conditions, a direct counter example can be given for $\rho>1$ and $M=1$ \cite{Hazer-WiOpt}.

However, our next theorem is promising for Rayleigh fading channels, and it shows that, homogenous feedback policies are optimal for a broad range of model parameters.
\begin{theorem} \label{Shur-concavity of Rate - MB}
For a Rayleigh fading environment, $R\paren{\vec{p}}$ is a Schur-concave function of feedback probabilities, and therefore a homogenous threshold feedback policy solves \eqref{Optimization Problem} if $M > 1$.
\end{theorem}

Both theorems are direct applications of Theorem \ref{Thm: Shur-concavity of Rate}.  Since the multi-beam scenario is a more general case, one may a priori assume that homogenous feedback policies should always be optimal for $M=1$ since they are optimal for $M>1$.  However, this intuition is not correct, and homogenous threshold optimality for the single beam case depends on the background noise level.  Some more practical insights on this result can be found in \cite{Hazer-WiOpt}.

\section{Conclusions} \label{Conclusions}
In this paper, we have analyzed optimal threshold selection problem for threshold feedback policies for finite feedback systems.  This is a non-convex optimization problem.  We have established sufficient conditions under which the rate function becomes a Schur-concave function by utilizing the theory of majorization.  Schur-concavity of the rate function automatically implies the optimality of homogenous threshold feedback policies in which all users use the same threshold value for their feedback decisions.  We illustrate an example application of our results to Rayleigh fading environments.  More refined sufficient conditions on the number of beams and the background noise level for the homogenous threshold optimality have been obtained in this case.  It is also seen that homogenous threshold feedback policies may not be optimal for a single beam system with low background noise levels.

\bibliography{Tharaka-bibfile}
\end{document}